\documentclass[a4paper, 12pt]{article}
\usepackage[english]{babel}
\usepackage[a4paper, inner=2cm, outer=2cm, top=3cm, bottom=3cm]{geometry}
\usepackage[dvipsnames]{xcolor}
\usepackage{mathtools}
\usepackage{pstricks}
\usepackage{caption}
\usepackage{graphicx}
\usepackage{amsmath}
\usepackage{amssymb}
\usepackage{mathcomp}
\usepackage{textcomp}
\usepackage{enumitem}
\usepackage{physics}
\usepackage{romannum}
\usepackage[doublespacing]{setspace}
\usepackage{subcaption}
\usepackage{hyperref}
\usepackage{xcolor}
\usepackage{booktabs}
\usepackage{cite}

\hypersetup{
  colorlinks   = true,
  urlcolor     = Blue, 
  linkcolor    = Black,
  citecolor    = Black
}

\renewcommand{\thesection}{\Roman{section}} 

\newcommand{\head}[1]{\textnormal{\textbf{#1}}}

\title{On the Gravitational Precession Memory Effect for an Ensemble of Gyroscopes}
\author{Raihaneh Moti$^1$ and Ali Shojai$^2$\footnote{Corresponding author: ashojai@ut.ac.ir}\\
\textit{{\small $^1$ School of Astronomy, Institute for Research in Fundamental Sciences (IPM), Tehran, Iran}} \\ \textit{{\small $^2$ Department of Physics, University of Tehran, Tehran, Iran}}}
\date{}

\begin{document}
\maketitle
\pagenumbering{arabic}
\begin{abstract}
We study the thermodynamic properties of a freely falling ensemble of gyroscopes after the passage of a weak gravitational wave. Due to the precession memory effect, the thermodynamic quantities will experience a change because of the space-time perturbation. We discuss that this \textit{GravoThermo memory effect} potentially can be used for the detection of the gravitational waves. 
\end{abstract}

%%%%%%%%%%%%%%%%%%%%%%%%%%%%%%%%%%%% Section I
\section{Introduction} \label{SecI}
The gravitational memory effect is a phenomenon caused by the gravitational disturbance of the background. It affects physical configurations, such as a group of particles with specific initial conditions. This effect is introduced in the literature of gravity theory through works of Zel'dovich and Polnaver \cite{Zeldovich, Smarr, Bontz, Garfinkle} and expanded by Christodoulou \cite{Christodoulou, Thorne} and others. 
The passage of a gravitational wave causes permanent changes in the physical configuration in the space-time and brings a variety of observable phenomena to the scene. The displacement memory effect \cite{Favata, Zhang}, the most known one, emerges as a net displacement between freely falling test particles in the wave zone. There are other effects such as kick memory \cite{Velocity1, Velocity2}, spin memory \cite{Pasterski, Nichols-Spin, Jokela}, center-of-mass memory \cite{Nichols-CM}, and gyroscopic memory \cite{Herrera, Seraj1, Seraj2} appearing in different contexts.

An important memory effect can be observed for a spinning object. A freely falling spinning gyroscope would experience precession with respect to the distant stars when a localized pulse of gravitational wave crosses its path.  An observer equipped with a gyroscope would measure a net change of orientation around its axis  just after the passage of the gravitational wave. If the gyroscope is suspended in a torque-free manner in a  freely falling frame, this precession, originating from the space-time perturbation, makes a distinction between before and after of the exposure to the gravitational pulse. This is the so-called gyroscopic memory and this is the one in which we are interested in, here.

The spin 4-vector $S^{\mu}$ of a small freely falling gyroscope is parallel transported along its worldline with proper velocity $v^{\mu}$ obeying the 
\begin{equation}
v^{\nu} ~ \nabla_{\nu} S^{\mu} = 0 \ .
\end{equation}
In a local frame  co-moving (defined by the fixed-stars oriented tetrads $e^{\hat{a}}_{~ \mu}$) with the gyroscope, for which $e^{~ \mu}_{\hat{0}} = v^{\mu}$, the gyroscope spin is purely spatial $S^{\hat{i}} = S^{\mu} e^{\hat{i}}_{~ \mu}$ and $S^{\hat{0}} = 0$. This choice best describes the spatial precession of the geodesics with respect to the fixed-stars.

The parallel transport equation then reads
\begin{equation}
\dv{S^{\hat{i}}}{s} = \Omega^{\hat{i}}_{~ \hat{j}} (s) S^{\hat{j}} (s)
\end{equation}
where $s$ is the observer's proper time and the angular velocity \cite{MTW}
\begin{equation}
\Omega^{\hat{i}\hat{j}} = - v^{\alpha} \omega_{\alpha}^{~\hat{i}\hat{j}}
\end{equation}
is the projection of the spin connections of the space-time $\omega_{\mu}^{~ \hat{\mu}\hat{\nu}} = e^{\hat{\mu}}_{~ \alpha}\nabla_{\mu}e^{~ \hat{\nu}\alpha}$ along $v^{\alpha}$.
Therefore, the gyroscope precession rate is determined by the spin connections evaluated along the observer's path. 

Because the gravitational disturbance of the background can be accurately described by the Bondi coordinate system \cite{BS formalism}, it is the suitable choice for coordinating the overall framework of this scenario. 
The retarded Bondi coordinates $x^{\alpha} = (u,r, \chi^{1},\chi^{2})$ are based on a family of outgoing null hypersurfaces $u=const$ (where  $u \equiv t-r$ is the retarded time), $r$ which varies along the null rays and the angular coordinates $\chi^{A}=(\theta,\phi)$ which are constant along the null rays. Near the future null infinity, the metric has the form 
\begin{equation}
\dd{s}^2 = -Ue^{2\tilde{\beta}} \dd{u^2} - 2e^{2\tilde{\beta}} \dd{u} \dd{r} +r^2\gamma_{AB}\left(\dd{\chi^A}-\mathcal{U}^A\dd{u}\right)\left(\dd{\chi^B}-\mathcal{U}^B\dd{u}\right)
\end{equation}
where $A,B=1,2$ and $U, \tilde{\beta}, \mathcal{U}^{A}$ and $\gamma_{AB}$ are functions of coordinates $(u,r,\chi^A)$. These functions are introduced in details in the \ref{App. A}.
At null hypersurface at infinity (i.e. the celestial sphere), source oriented local frame is defined as the one with its time-like vector along the retarded time $u$. Through a spatial rotation, this local frame can be tied to the distant stars \cite{Seraj1, Seraj2}.

Using these considerations it is shown \cite{Seraj1,Seraj2,Seraj3} that as an effect of the gravitational wave passage, a gyroscope initially oriented in an arbitrary direction, $S_{\hat{i}}(u_0)$, would experience (up to $\mathcal{O}(r^{-2})$) a net precession 
\begin{equation}
\Delta S^{\hat{r}} = \mathcal{O}(r^{-3}) \quad \quad, \quad \quad \Delta S^{\hat{A}} = \epsilon^{\hat{A}\hat{B}} S_{\hat{B}}(u_0) \Delta\Phi + \mathcal{O}(r^{-3})
\label{Precession}
\end{equation}
where
\begin{equation}
\Delta\Phi =-\dfrac{1}{r^2} \int \dd{u} \left(\dfrac{1}{4} \nabla_A \nabla_B \tilde{C}^{AB} - \dfrac{1}{8} N_{AB} \tilde{C}^{AB} \right) \ .
\label{memory}
\end{equation}
$\tilde{C}_{AB}\equiv \epsilon_{AC} C_{B}^{\ C}$ is \textit{dual Bondi shear}, and the \textit{Bondi shear}, $C_{AB}$, is defined as the shear of a congruence of outgoing null rays which encodes the gravitational waveform received at the retarded time $u$ and the angles $\chi^A$ on the celestial sphere. The \textit{news tensor} $N_{AB}\equiv \partial_u C_{AB}$ determines time dependence of the Bondi shear and is a measure of the energy flux of the gravitational radiation. The first term of $\Delta\Phi$ is similar to what appears in the spin memory effect, given by the difference $\Delta T$ in travel time $T$ of the two beams, and is the effect of gravitational waves on a Sagnac interferometer \cite{Pasterski}. The second nonlinear term is related to the charge of gravitational electric--magnetic duality, as shown in \cite{Seraj1,Seraj2}.
Any index with a hat represent the index in the local frame.

Such a gyroscopic memory effect could in principle be the basis of a gravitational wave detector and this is what we are dealing with in this paper. To see how this is possible, here we consider an ensemble of charged gyroscopes aligned randomly in the presence of a magnetic field and located in a freely falling local box far from the  gravitational wave source. Assuming that the source of the gravitational wave is submitting a pulse wave, the world-line of the box would be disturbed as the pulse passes the box location. 
In addition to this, each member of the ensemble would change its spin direction after the passage of the gravitational pulse according to the equations \eqref{Precession} and \eqref{memory}. This would lead to a change in the Hamiltonian of the system and thus a change in the  partition function and in the thermodynamic properties of the system. Therefore, any gravitational wave can be detected by the corresponding change of the thermodynamic quantities of this system.

As an example consider that our freely falling box is filled with a (classical) paramagnetic matter. If this box is also in a local magnetic field, then one expects that the thermodynamic properties like entropy of the system would change after the exposure to the gravitational pulse. As a result such a system can \textit{sense} the passage of gravitational waves in a thermodynamic manner, an effect which can be called \textit{GravoThermo} memory effect.

This paper is organized as follows. First in the section \ref{SecII}, we construct the model for thermodynamic relations of a freely falling box of classical paramagnetic matter in the presence of a local magnetic field. Then, the response of such a system to the passage of a gravitational wave will be studied in section \ref{SecIII}. We will see how such a system can detect both the direction and magnitude of a gravitational wave source.

%%%%%%%%%%%%%%%%%%%%%%%%%%%%%%%%%%%% Section II
\section{Classical Paramagnetic Matter and the Gravitational Precession Memory Effect} \label{SecII}
\begin{figure}
\begin{center}
\includegraphics[scale=0.75]{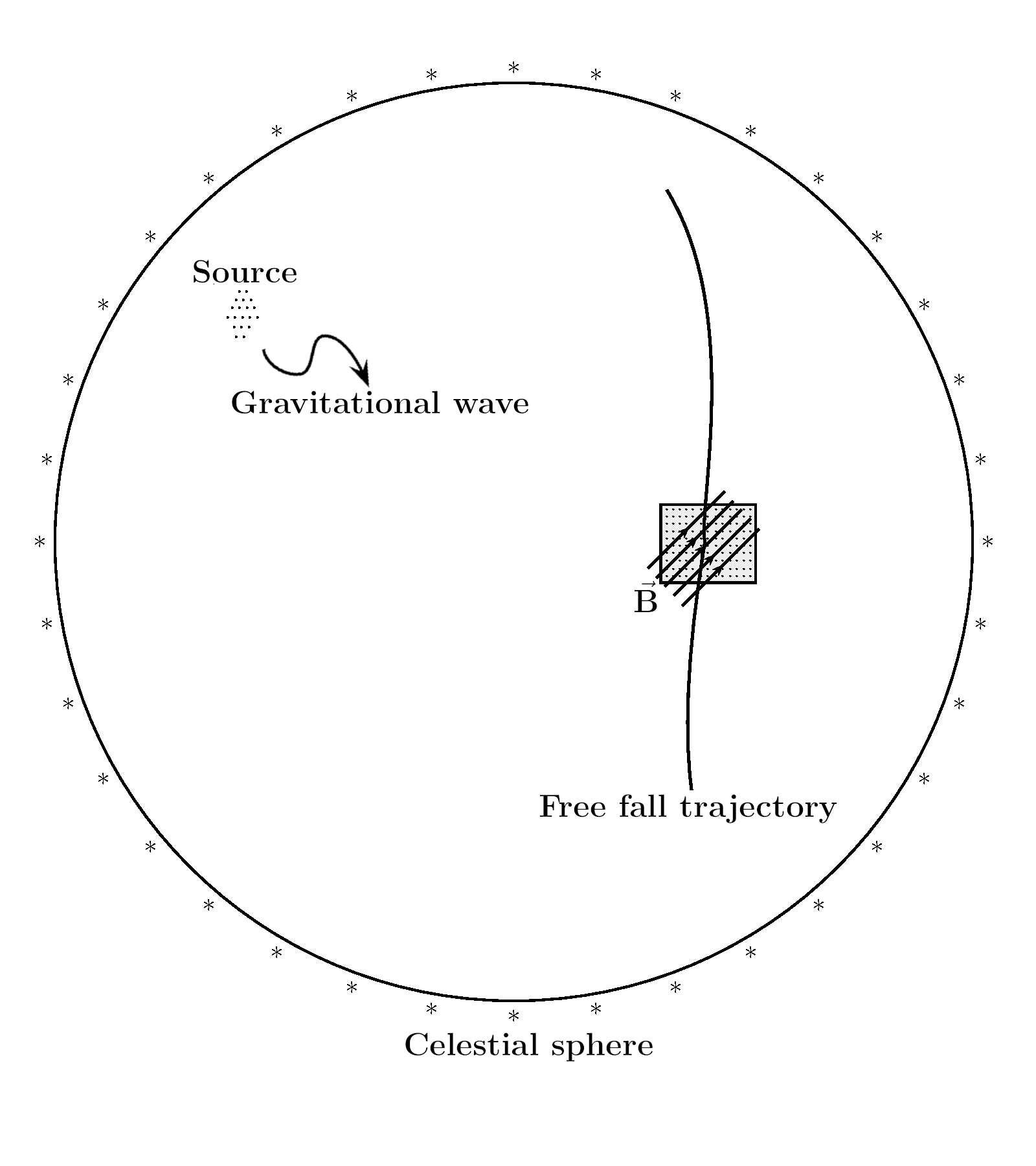}
\caption{A small freely falling box of paramagnetic matter in the presence of a local magnetic field. A distant source of gravitational wave sends a pulse towards the box. Local inertial frame basis are made non-rotating with respect to the distant galaxies seen on the celestial sphere.}
\label{theFIG}
\end{center}
\end{figure}

As it was stated in the Introduction, consider a small freely falling box of classical paramagnetic matter in the presence of a local magnetic field, as it is shown in Fig. (\ref{theFIG}). A distant source sends a gravitational wave pulse towards the box. One can adopt the Bondi coordinate system and thus the pure precession of each spinning member of the paramagnetic matter is measured with respect to the distant galaxies, i.e. the celestial sphere.

In addition we can choose the box to be enough small so that the neighboring gravitational interaction between gyroscopes (i.e. different members of the matter inside the box) can be forgotten. In other words we are neglecting the tidal force within the box. 

Before arrival of the wave pulse, at some Bondi retarded time $u_0$, the partition function of the system can be derived as \cite{Pathria}
\begin{equation}
\mathcal{Z}^{(0)} = \int e^{-\beta H_0} \dd{\Omega}
\end{equation}
where $H_0 = -\mu \vec{B}\cdot\vec{S}_0$, $\mu$ is the magnetic moment of each atom in the box and $\vec{S}_0\equiv \vec{S}(u_0)$. 
The described model is applicable in the semi-classical limit, where $H_0$ represents the Hamiltonian governing the classical interaction between the magnetic field and the spin. It is important to note that this model is a close approximation of the gyroscopes ensemble, where we ignore any interaction between neighboring spins. Moreover, the volume of the ensemble is fixed and we are dealing with a canonical ensemble. Although we are considering here a canonical ensemble of classical paramagnetic matter, the same reasoning makes classical or quantum magnetized matter either in canonical or grand canonical ensembles, as a suitable detector for gravitational waves. This is because of the fact that the gravitational wave changes the partition function as it passes the matter, and as the system has memory any thermodynamic measurement shows a change as a signal for passage of the wave.

As the gravitational wave pulse passes the box, freely falling gyroscopes would experience a precession according to Eq. \eqref{memory}. Therefore at some time $u$ (which is after the passage of pulse), the memory effect causes each spin to be equal to $\vec{S} =\vec{S}_0 +\Delta \vec{S}$. The partition function would be now
\begin{equation}
\mathcal{Z} = \int e^{-\beta\left (-\mu \vec{B}\cdot(\vec{S}_0 +\Delta \vec{S})\right )} \dd{\Omega} \ .
\end{equation}

Without loss of generality we assume that the magnetic field is orthogonal to the $r$ coordinate,  and show its orientation in the $\chi^A$ surface with respect to the spin by the angle $\vartheta$. On the other hand, by Eq. \eqref{Precession}, for each gyroscope the projection of $\Delta \vec{S}$ is perpendicular to the projection of $\vec{S}_0 $ onto the 2-surface and thus  $\Delta \vec{S}$ orientation with respect to the magnetic field is $\vartheta'\equiv -\vartheta+\dfrac{\pi}{2}$. Thus, for  $ \abs{B} = const.$ , 
\begin{equation}
\mathcal{Z} =  \int e^{\beta\mu B S_0 \cos{\vartheta}+  \beta\mu B \Delta S \cos{\vartheta'}} \dd{\Omega} 
\end{equation}
where $\Delta S=S_0 \abs{\Delta \Phi}$, and $S_0$ and $\Delta S$ are the length of $\vec{S}_0$ and $\Delta\vec{S}$ projected onto the 2-surface, respectively. Therefore
\begin{equation}
\mathcal{Z} =  2\pi \int e^{\beta\mu B S_0 \cos{\vartheta}}\ \times  e^{\beta\mu B S_0 \Delta\Phi \sin{\vartheta}}  \dd{(\cos{\vartheta})} \ .
\end{equation}
Since $\Delta\Phi \propto -1/r^2$,  far from the source we can use the approximate relation
\begin{equation}
\mathcal{Z}  \simeq 2\pi \int e^{\beta\chi \cos{\vartheta}}  \bigl(1+  \beta\chi\Delta\Phi \sin{\vartheta}  \bigr) \dd{(\cos{\vartheta})} \equiv \mathcal{Z}^{(0)}+\Delta\Phi \tilde{\mathcal{Z}}
\label{PF}
\end{equation}
in which $ \chi \equiv \mu B S_0$. 

This is an important relation as it shows clearly how the memory effect footprint contributes in the partition function.
As a result, as long as $\Delta\Phi$ is small (which is correct for a distant gravitational source), all the thermodynamic quantities can be written as the sum of two terms; a no-memory term and a memory term which is proportional to $\Delta\Phi$, See Table \ref{Table1}. The quantities with a tilde sign are the contributions from gyroscope precession.

\begin{table}
\centering
\begin{tabular}{lrl}
\toprule[1.5pt]
\head{Thermodynamic Quantity}  & & \head{Expression}  \\
\midrule\\
Specific Heat ($\mathcal{C}_v$) &  $  k_B \beta^2 \pdv[2]{\ln{\cal Z}}{\beta}$ & $\equiv \mathcal{C}_v^{(0)}+\Delta\Phi \tilde{\mathcal{C}}_v$  \\[2ex]
Entropy ($\mathcal{S}$) &  $k_B \ln{\mathcal{Z}} - k_B \beta \pdv{\ln{\mathcal{Z}}}{\beta} $ & $\equiv \mathcal{S}^{(0)}+\Delta\Phi \tilde{\mathcal{S}}$  \\[2ex]
Average Energy ($\mathcal{E}$) & $-\pdv{\ln{\mathcal{Z}}}{\beta}$ & $\equiv \mathcal{E}^{(0)}+\Delta\Phi \tilde{\mathcal{E}}$  \\[2ex]
Helmholtz Free Energy (${\cal F}$)  &  $-\frac{1}{\beta} \ln{\mathcal{Z}}$ & $\equiv \mathcal{F}^{(0)}+\Delta\Phi \tilde{\mathcal{F}}$  \\[2ex]
Gibbs Free Energy (${\cal G}$) &  $ -\frac{1}{\beta}\ln{\mathcal{Z}} + \frac{V}{\beta} \pdv{\ln{\mathcal{Z}}}{V}$ & $\equiv \mathcal{G}^{(0)}+\Delta\Phi \tilde{\mathcal{G}}$  \\[2ex]
Enthalpy ($ {\cal H}$) & $   - \frac{\partial\ln{\mathcal{Z}}}{\partial\beta} + \frac{V}{\beta}\pdv{\ln{\mathcal{Z}}}{V} $ & $\equiv \mathcal{H}^{(0)}+\Delta\Phi \tilde{\mathcal{H}} $ \\[2ex]
\bottomrule[1.5pt]
\end{tabular}
\caption{All the thermodynamic quantities are the sum of a no-memory term and a memory term which is proportional to $\Delta\Phi$. }
\label{Table1}
\end{table}

%%%%%%%%%%%%%%%%%%%%%%%%%%%%%%%%%%%% Section III
\section{Response to the Passage of a Gravitational Wave}\label{SecIII}
In order to see how the memory effect affects the thermodynamic properties of the matter in the falling box after the passage of a gravitational pulse, the Bondi shear of the space-time at null infinity defined by \eqref{memory} is required. It is clear that the source properties are encoded into this tensor \cite{Seraj4}. To make the results as general as possible and more independent of the source properties, we use some general assumptions. As one of the most prominent properties of the Bondi space-time is its \textit{asymptotic flatness}, far enough from the source the gravitational wave can be viewed as a perturbation to the flat space-time. 

Therefore one can consider a general gravitational pulse with both the polarization directions $h_+$ and $h_{\times}$ in the asymptotic flat space-time with a metric 
\begin{equation}
g_{\mu\nu}^{\text{(Cartesian)}}=\eta_{\mu\nu}+h_{\mu\nu} =
\begin{pmatrix}
-1 & 0 & 0 & 0 \\
0 & 1+h_+ ~ f(u,r) & h_{\times} ~ f(u,r)  & 0 \\
0 & h_{\times} ~ f(u,r) & 1 -h_+ ~ f(u,r) & 0 \\
0 & 0 & 0 & 1
\end{pmatrix}
\end{equation}
in the Cartesian coordinates $(t,x,y,z)$. Here $f(u,r)$ is a localized function of the retarded time and radius and describes the pulse and its passage through the box.
In the null spherical coordinates $(u,r,\theta,\phi)$, this perturbed space-time would be described by
\begin{footnotesize}
\begin{equation}
\begin{aligned}
g_{\mu\nu} &  = \\
& = 
\left(\begin{matrix}
-1 & 1  \\
1 & \sin^2\theta \left(\tilde{h}_+ \cos2\phi + \tilde{h}_{\times}\sin2\phi \right)  \\
0 &  \dfrac{r}{2}\sin2\theta\left(\tilde{h}_+ \cos2\phi + \tilde{h}_\times \sin2\phi \right) \\
0 & r\sin^2\theta \left(\tilde{h}_+\cos2\phi - \tilde{h}_+\sin2\phi\right) 
\end{matrix}
\right.
\\
\\
&  \qquad \qquad \left.
\begin{matrix}
0 & 0 \\
 \dfrac{r}{2}\sin2\theta\left( \tilde{h}_+\cos2\phi + \tilde{h}_\times \sin2\phi \right)& 
r\sin^2\theta\left(\tilde{h}_+\cos2\phi - \tilde{h}_+\sin2\phi\right) \\
 r^2 + r^2 \cos^2\theta \left(\tilde{h}_+ \cos2\phi + \tilde{h}_\times\sin2\phi \right)  & \dfrac{r^2}{2}\sin2\theta\left(\tilde{h}_\times\cos2\phi - \tilde{h}_+\sin2\phi\right)\\
 \dfrac{r^2}{2}\sin2\theta\left(\tilde{h}_\times\cos2\phi - \tilde{h}_+\sin2\phi\right) & -r^2\sin^2\theta\left(\tilde{h}_+ \cos2\phi + \tilde{h}_\times\sin2\phi\right) + r^2\sin^2\theta
\end{matrix} \right)
\end{aligned}
\end{equation}
\end{footnotesize}
where $ \tilde{h}_{+/\times} = h_{+/\times} ~ f(u,r)$.

The polarization amplitudes $h_+$ and $h_\times$ have to be calculated using the linearized gravity in terms of the second mass moments, $M_{ij}$, as it can be found in any textbook on gravitational waves, see e.g. \cite{Maggiore}. The result is

\begin{align}
 \tilde{h}_+ = &  \ddot{M}_{11} \left(\cos^2\phi-\sin^2\phi \cos^2\theta\right)+ \ddot{M}_{22} \left(\sin^2\phi-\cos^2\phi\cos^2\theta\right) \nonumber \\ 
 & -\ddot{M}_{33}\sin^2\theta -\ddot{M}_{12}\sin2\phi \left(1+\cos^2\theta\right) + \ddot{M}_{13}\sin\phi\sin2\theta + \ddot{M}_{23} \cos\phi\sin2\theta  \label{A11} \\
\tilde{h}_\times = & \left(  \ddot{M}_{11} -\ddot{M}_{22} \right) \sin2\phi\cos\theta + 2\ddot{M}_{12}\cos2\phi \cos\theta -2 \ddot{M}_{13}\cos\phi\sin\theta + 2 \ddot{M}_{23} \sin\phi\sin\theta \ . \label{A22}
\end{align}
A dot over any quantity shows derivative with respect to the $t$.

In order to calculate the Bondi shear and thus the precession factor $\Delta\Phi$, one needs the 2-metric of $\chi^A$ surface
\begin{footnotesize}
\begin{equation}
\xi_{AB} =
\begin{pmatrix}
 r^2 + r^2 \cos^2\theta \left(\tilde{h}_+ \cos2\phi + \tilde{h}_\times\sin2\phi \right) & \dfrac{r^2}{2}\sin2\theta\left(\tilde{h}_\times\cos2\phi - \tilde{h}_+\sin2\phi\right)  \\
 \dfrac{r^2}{2}\sin2\theta\left(\tilde{h}_\times\cos2\phi - \tilde{h}_+\sin2\phi\right)  & -r^2\sin^2\theta\left(\tilde{h}_+ \cos2\phi + \tilde{h}_\times\sin2\phi\right) + r^2\sin^2\theta
\end{pmatrix} \ .
\end{equation}
\end{footnotesize}

On using the Eq. \eqref{Shear} of the \ref{App. A}, one gets the relation
\begin{small}
\begin{equation}
C_{AB} = 
\begin{pmatrix}
 \cos^2\theta \left(\tilde{h}_+ \cos2\phi + \tilde{h}_\times\sin2\phi \right)
 & \dfrac{1}{2}\sin2\theta\left(\tilde{h}_\times\cos2\phi - \tilde{h}_+\sin2\phi\right) \\
 \dfrac{1}{2}\sin2\theta\left(\tilde{h}_\times\cos2\phi - \tilde{h}_+\sin2\phi\right)  & -\sin^2\theta\left(\tilde{h}_+ \cos2\phi + \tilde{h}_\times\sin2\phi\right)
\end{pmatrix}
\label{flat shear}
\end{equation}
\end{small}
for Bondi shear and 
\begin{scriptsize}
\begin{align}
\tilde{C}^{\theta\theta} & =  \dfrac{ \tilde{h}_\times\cos2\phi - \tilde{h}_+\sin2\phi  }{\left(-2 +( \tilde{h}_+^2 + \tilde{h}_\times^2)(1+\cos2\theta) +2\sin^2\theta (\tilde{h}_+\cos2\phi + \tilde{h}_\times\sin2\phi)  \right)^2} 4 \cot\theta \label{ZYX}\\
\tilde{C}^{\theta\phi} & = 
 \dfrac{ ( \tilde{h}_+^2 + \tilde{h}_\times^2) (-3+\cos4\theta) - (\tilde{h}_+\cos2\phi + \tilde{h}_\times \sin2\phi)(9+\cos4\theta) -2\cos2\theta (2 + \tilde{h}_+^2 + \tilde{h}_\times^2   - \tilde{h}_+\cos2\phi - \tilde{h}_\times \sin2\phi) }{8 \left( \tilde{h}_+^2 + \tilde{h}_\times^2 - \tilde{h}_+\cos2\phi - \tilde{h}_\times \sin2\phi -(-1 + \tilde{h}_+^2 + \tilde{h}_\times^2 ) \csc^2\theta\right)^2 } \csc^6\theta\\
\tilde{C}^{\phi\theta} & =  \dfrac{2\cos2\theta (2 + \tilde{h}_+^2 + \tilde{h}_\times^2 - 3\tilde{h}_+\cos2\phi - 3\tilde{h}_\times \sin2\phi ) + (-3+\cos4\theta)(-\tilde{h}_+^2 - \tilde{h}_\times^2 + \tilde{h}_+\cos2\phi+ \tilde{h}_\times \sin2\phi ) }{8 \left( \tilde{h}_+^2 + \tilde{h}_\times^2 - \tilde{h}_+\cos2\phi - \tilde{h}_\times \sin2\phi -(-1 + \tilde{h}_+^2 + \tilde{h}_\times^2 ) \csc^2\theta\right)^2}  \csc^6\theta\\
\tilde{C}^{\phi\phi} & = \dfrac{-\tilde{h}_\times \cos2\phi + \tilde{h}_+\sin2\phi  }{ \left( \tilde{h}_\times^2 + \tilde{h}_+^2 - \tilde{h}_+\cos2\phi - \tilde{h}_\times\sin2\phi + (1 - \tilde{h}_\times^2 - \tilde{h}_+^2)\csc^2\theta \right)^2 } \cot\theta\csc^6\theta 
\end{align}
\end{scriptsize}
for the dual Bondi shear $\tilde{C}_{AB}\equiv \epsilon_{AC} C_{B}^{\ C}$.

To calculate $\Delta\Phi$, also the components of the news tensor $N_{AB}$ are needed, which can be obtained for \eqref{flat shear} as
\begin{align}
&  N_{\theta\theta} = 
 \left(\tilde{h}_+' \cos2\phi + \tilde{h}_{\times}'\sin2\phi \right) \cos^2\theta \\
&  N_{\theta\phi} =   N_{\phi \theta} = \dfrac{1}{2}\left(\tilde{h}_{\times}'\cos2\phi - \tilde{h}_+'\sin2\phi \right) \sin2\theta\\
 & N_{\phi\phi} = -\left(\tilde{h}_+' \cos2\phi + \tilde{h}_{\times}' \sin2\phi \right) \sin^2\theta \label{XYZ}
\end{align} 
where a prime over any quantity shows differentiation with respect to $u$.

In evaluating $\Delta\Phi$ from Eq. \eqref{memory}, one have to note that since the falling box of matter is considered small, we can assume $\theta$ and $\phi$ constant and the time dependence of the second mass moments are averaged over time because of the integral over $u$. As a result the quantity $\Delta\Phi$ can be taken out of the integral when calculating the partition function. The explicit form of $\Delta\Phi$ as a function of the second mass moments and angles $(\theta,\phi)$ is too lengthy to be written here. Equations \eqref{memory}, (\ref{A11}-\ref{A22}), and (\ref{ZYX}-\ref{XYZ}) completely determine the dependence of $\Delta\Phi$ on the direction of the incoming gravitational pulse, $(\theta,\phi)$, and its magnitude through $M_{ij}$.

The partition function, then can be simply calculated as
\begin{align}
\mathcal{Z}   & =    2\pi \int e^{\beta\chi \cos{\vartheta}}  \bigl(1+  \beta\chi\Delta\Phi \sin{\vartheta}  \bigr) \dd{(\cos{\vartheta})} \nonumber \\
& =\mathcal{Z}^{(0)} +2 \pi^2\Delta\Phi  I_1(\beta\chi)
\end{align}
where $I_{1}(x)$ is the modified Bessel function (of rank 1) of first kind and ${\cal Z}^{(0)}=\dfrac{2\sinh \beta\chi}{\beta\chi}  $ is the partition function with no memory effect.

Having the partition function at hand, all the thermodynamic quantities of the box of paramagnetic matter can be derived. The important ones are:

\begin{itemize}
\item Entropy (${\cal S}$):
\begin{multline}
{\cal S}=k_B\left(\ln\left(\frac{2 \sinh\beta\chi}{\beta\chi}+2 \pi^2\Delta\Phi I_1(\beta\chi)\right)+\right.\\
 \left.\frac{2(\sinh\beta\chi-\beta\chi\cosh\beta\chi)-\pi^2\Delta\Phi(\beta\chi)^2(I_0(\beta\chi)+I_2(\beta\chi))}{2\sinh\beta\chi+2\pi^2\Delta\Phi\beta\chi I_1(\beta\chi)}\right) 
\end{multline}
which for $\abs{\Delta\Phi} \ll 1$, reads as
\begin{equation}
{\cal S} = {\cal S}^{(0)} + \dfrac{k_B \pi^2 (\beta\chi)^2 }{2} \csch \beta\chi  \Bigl( -I_0(\beta\chi)+ 2 I_1(\beta\chi) \coth \beta\chi -I_2(\beta\chi)  \Bigr)  \Delta\Phi \ .
\end{equation}
The entropy without memory term is
\begin{equation}
{\cal S}^{(0)}=k_B\left(1-\beta\chi\coth\beta\chi+\ln\left(\frac{2 \sinh\beta\chi}{\beta\chi}\right)\right)
\end{equation}
in agreement with the classical paramagnetic matter partition function.
\item Average Energy (${\cal E}$): 
\begin{equation}
{\cal E} = \dfrac{2\sinh\beta\chi -\beta\chi\left(\pi^2\beta\chi\Delta\Phi (I_0(\beta\chi) +I_2(\beta\chi) ) +2\cosh\beta\chi \right)  }{2\beta \left(\pi^2\beta\chi\Delta\Phi I_1(\beta\chi) +\sinh\beta\chi \right) } \ .
\end{equation}
For $\abs{\Delta\Phi} \ll 1$, we have
\begin{equation}
{\cal E}={\cal E}^{(0)}-\pi^2\beta\chi^2 \Delta\Phi \left(-I_0(\beta\chi)+I_1(\beta\chi)\coth\beta\chi \right)  \csch\beta\chi
\end{equation}
where ${\cal E}^{(0)}= \left(1-\beta\chi\coth\beta\chi \right)/\beta$ has the form expected.

\item Specific Heat (${\cal C}_v$):
\begin{multline}
 {\cal C}_v = \dfrac{k_B}{4 \left( \pi^2\beta\chi\Delta\Phi I_1(\beta\chi) + \sinh\beta\chi \right)^2 } \times \\
  \left(        
-\left(\pi^2\beta^2\chi^2\Delta\Phi \left(I_0(\beta\chi) +  I_2(\beta\chi) \right)  + 2\beta\chi\cosh \beta\chi - 2\sinh\beta\chi \right)^2 \right. \\
\left. + \left( \pi^2\beta\chi\Delta\Phi I_1(\beta\chi) + \sinh\beta\chi \right) \times \right. \\
\left. \left( 3\pi^2\beta^3\chi^3 \Delta\Phi (I_1(\beta\chi)+I_3(\beta\chi)) +8\sinh\beta\chi -8\beta\chi\cosh\beta\chi +4\beta^2\chi^2\sinh\beta\chi \right)
 \right) \ .
\end{multline}
This can be expanded for $\abs{\Delta\Phi} \ll 1$ as
\begin{multline}
 {\cal C}_v ={\cal C}_v^{(0)} + \dfrac{1}{2}k_B \pi^2 (\beta\chi)^2  \csch^3\beta\chi \times \\
  \Bigl( 4\beta\chi \cosh^2\beta\chi I_1(\beta\chi)  + I_0(\beta\chi) \left(-1 +\cosh2\beta\chi -2 \beta\chi\sinh2\beta\chi \right) \Bigr) \Delta\Phi
    \label{SH}
\end{multline}
where
\begin{equation}
{\cal C}_v^{(0)} = k_B \left( 1+(\beta\chi)^2 -(\beta\chi)^2\coth^2\beta\chi \right)
\end{equation}
is the specific heat of the paramagnetic matter without memory effect.
\end{itemize}

\begin{figure}[ht!]
\centering
\includegraphics[keepaspectratio=true, width=0.7\textwidth]{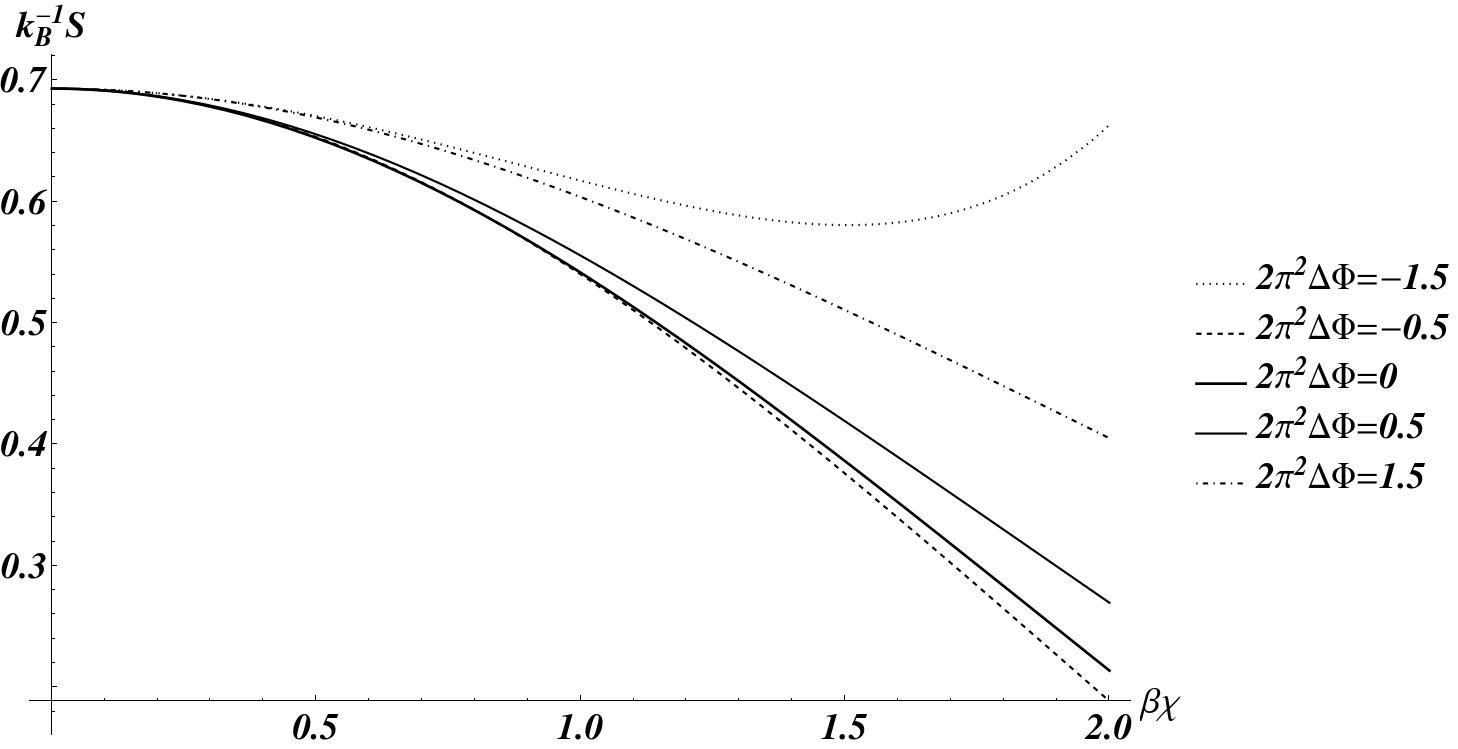}
\caption{Plot of Entropy for different values of $\Delta\Phi$}
 \label{Entropy-chi}
\end{figure}

\begin{figure}[ht!]
\centering
\includegraphics[keepaspectratio=true, width=0.7\textwidth]{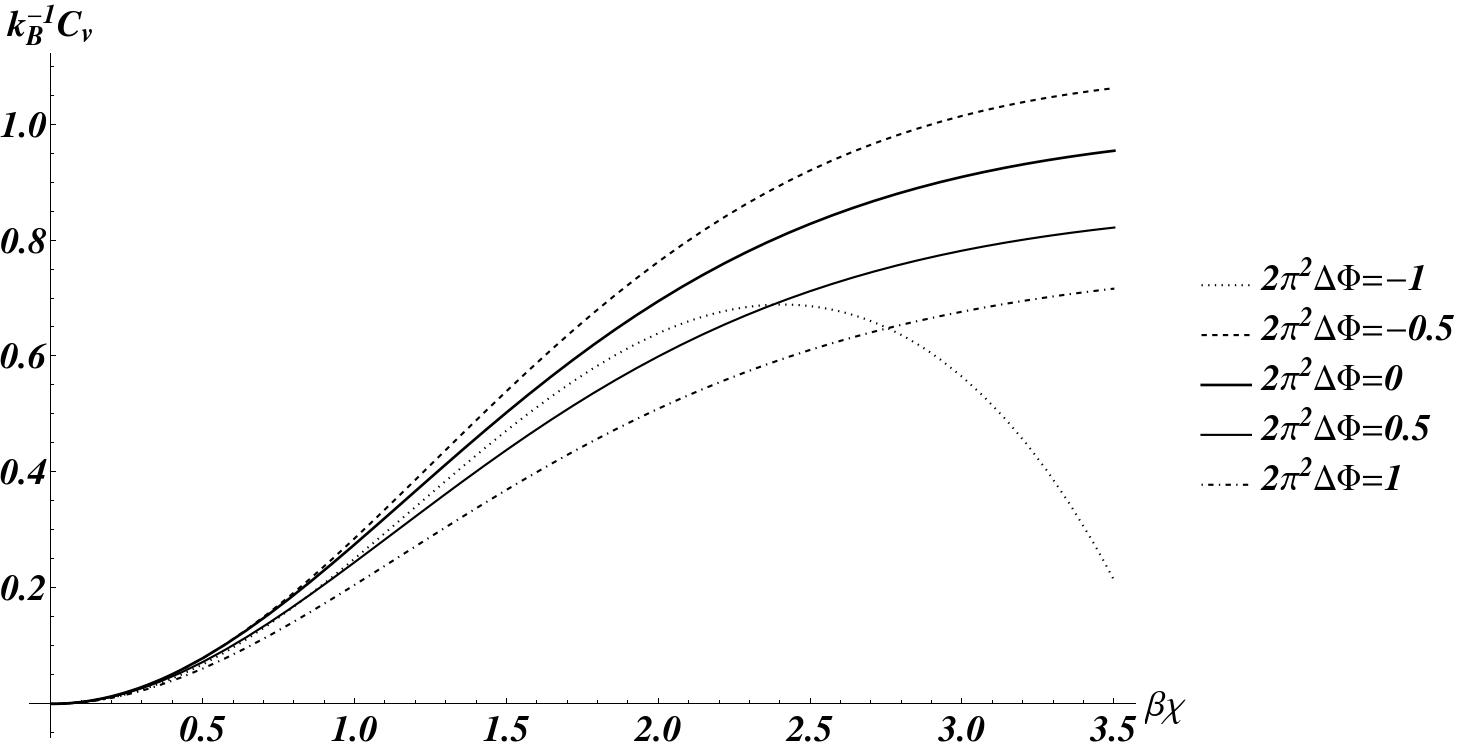}
\caption{Plot of Specific heat for different values of $\Delta\Phi$}
 \label{Specific heat-chi}
\end{figure}

In order to see how the memory effect changes the thermodynamic properties of our freely falling box of paramagnetic matter in a magnetic field, entropy and specific heat are compared for different values of $\Delta\Phi$ in figures (\ref{Entropy-chi}) and (\ref{Specific heat-chi}), respectively. 
It can be seen that for negative values of $\Delta\Phi$ both entropy and specific heat as a function of $\beta\chi$ change their slope at some value of $\beta\chi$. 

\begin{figure}
     \centering
     \begin{subfigure}[b]{0.45\textwidth}
         \centering
         \includegraphics[width=\textwidth]{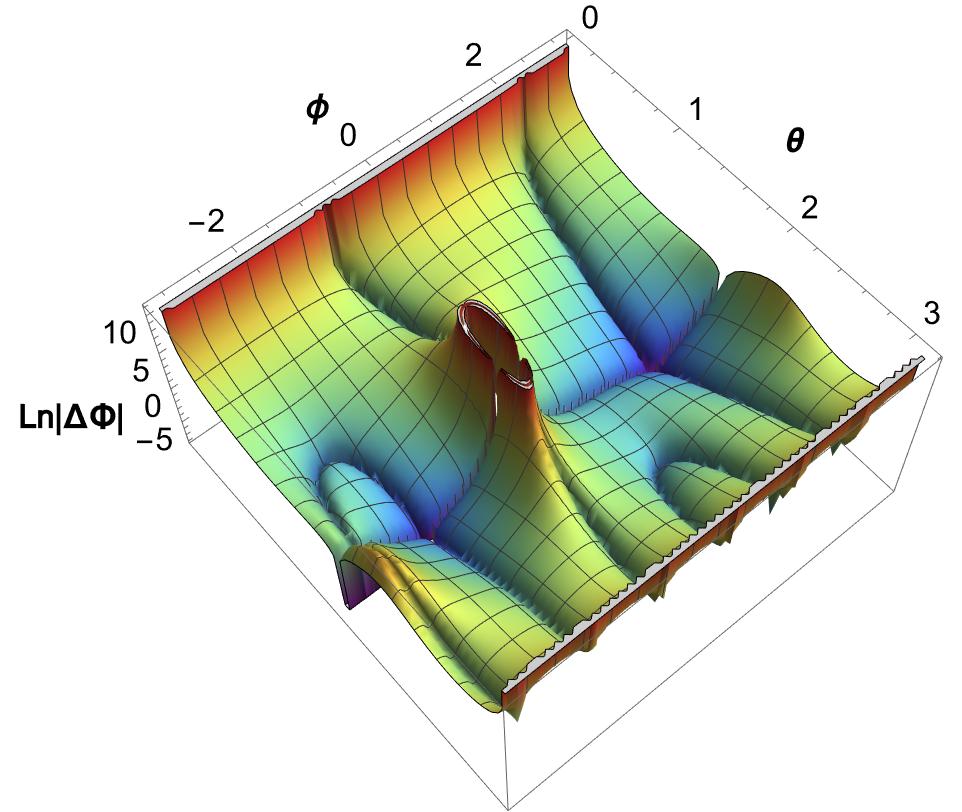}
         \caption{Case 1: $\overline{\ddot{M}_{13}}\sim \overline{\ddot{M}_{23}}\sim \overline{\ddot{M}_{33}} \ll \overline{\ddot{M}_{11}}\sim \overline{\ddot{M}_{22}}$}
         \label{deltaphi1}
     \end{subfigure}
     \hfill
     \begin{subfigure}[b]{0.45\textwidth}
         \centering
         \includegraphics[width=\textwidth]{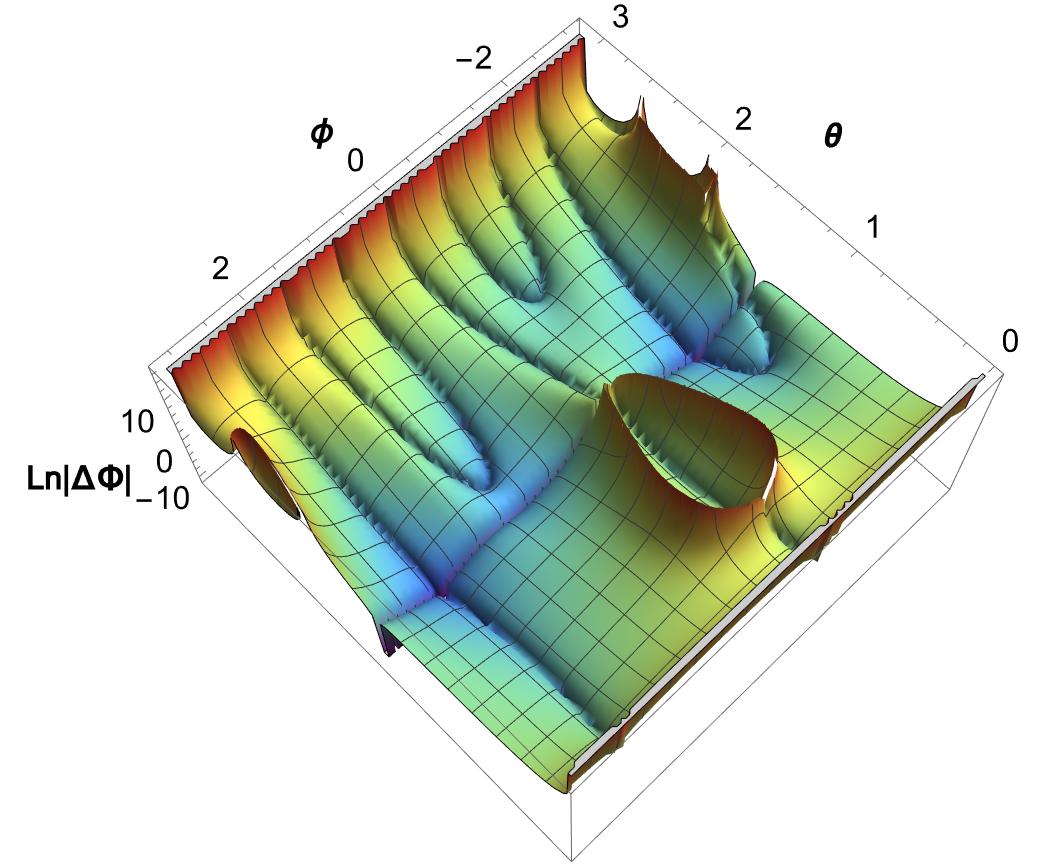}
         \caption{Case 2: $\overline{\ddot{M}_{13}}\sim \overline{\ddot{M}_{23}}\sim \overline{\ddot{M}_{33}} \sim \overline{\ddot{M}_{22}}\ll \overline{\ddot{M}_{11}}$}
         \label{deltaphi2}
     \end{subfigure}
        \caption{Dependence of normalized $\Delta\Phi$ (normalized by setting $\overline{\ddot{M}_{11}}=1$ and $r=1$) on the direction of incoming gravitational pulse.}
        \label{deltaphi}
\end{figure}

An interesting result is that as the value of $\Delta\Phi$ depends on values of $\theta$ and $\phi$, and as the thermodynamic properties (like entropy and specific heat) are dependent on $\Delta\Phi$, our falling box can detect the direction of the incoming gravitational pulse. To see this in a more clear way, let us first investigate the angular dependence of $\Delta\Phi$.  Actually this is also dependent on the details of the second mass moments. As an example consider an axial symmetric source of gravitation waves in two cases: first the case of $\overline{\ddot{M}_{13}}\sim \overline{\ddot{M}_{23}}\sim \overline{\ddot{M}_{33}} \ll \overline{\ddot{M}_{11}}\sim \overline{\ddot{M}_{22}}$ and similar relation for the third time derivatives of mass moment, and second the case of $\overline{\ddot{M}_{13}}\sim \overline{\ddot{M}_{23}}\sim \overline{\ddot{M}_{33}} \sim \overline{\ddot{M}_{22}}\ll \overline{\ddot{M}_{11}}$ and similar relation for the third time derivatives of mass moment (A bar over any quantity denotes its time average). Angular dependence of logarithm of the absolute value of normalized $\Delta\Phi$ (normalized by setting $\overline{\ddot{M}_{11}}=1$ and $r=1$) is shown in Fig. (\ref{deltaphi}).

As the direction of the wave changes, the value of $\Delta\Phi$ changes alternatively from negative values to positive ones and back to negative values. This will introduce alternate changes in the thermodynamic quantities.

\begin{figure}
     \centering
     \begin{subfigure}[b]{0.49\textwidth}
         \centering
         \includegraphics[width=\textwidth]{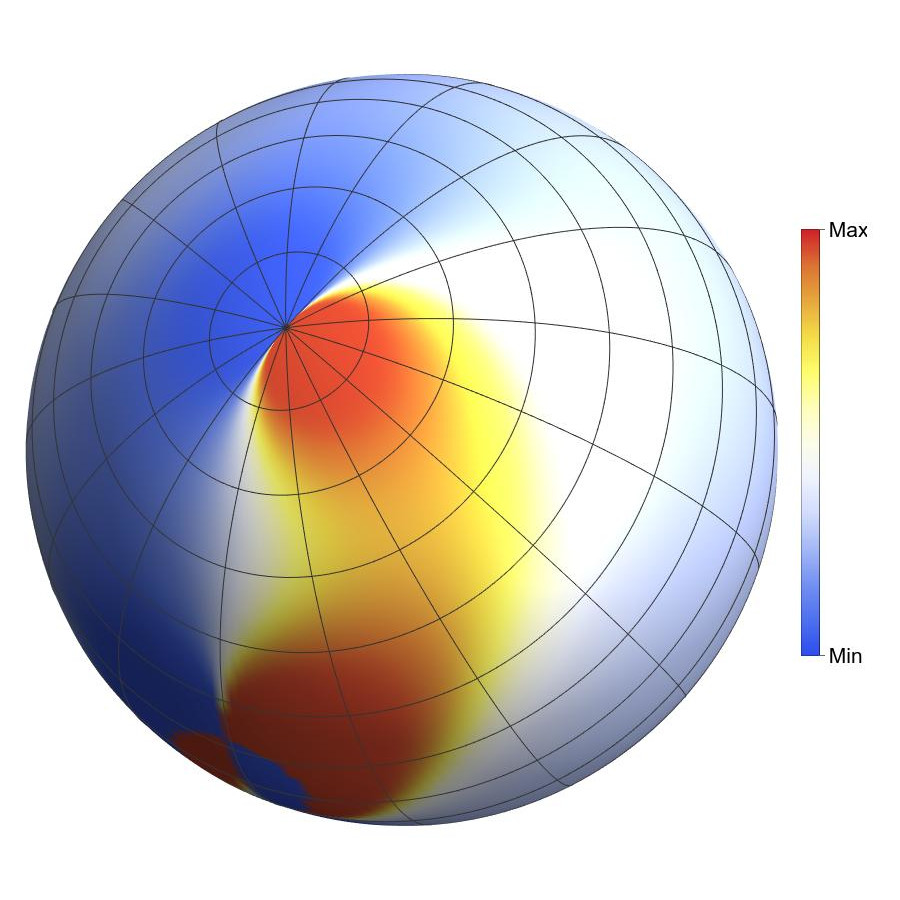}
         \caption{$\frac{\Delta {\cal C}_v}{{\cal C}_v}$, view from the north pole.}
         \label{cv1}
     \end{subfigure}
     \begin{subfigure}[b]{0.49\textwidth}
         \centering
         \includegraphics[width=\textwidth]{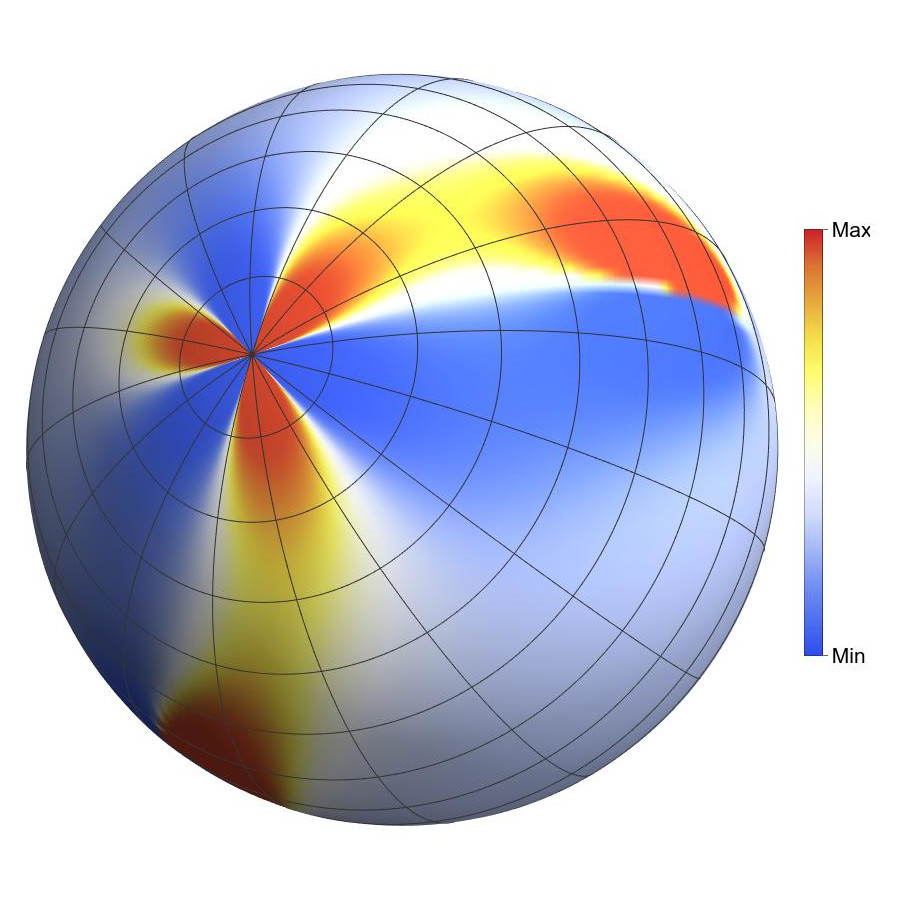}
         \caption{$\frac{\Delta {\cal C}_v}{{\cal C}_v}$, view from the south pole.}
         \label{cv1t}
         \end{subfigure}
        \caption{Dependence of $\frac{\Delta {\cal C}_v}{{\cal C}_v}$ on the direction of incoming gravitational pulse for the Case $\overline{\ddot{M}_{13}}\sim \overline{\ddot{M}_{23}}\sim \overline{\ddot{M}_{33}} \ll \overline{\ddot{M}_{11}}\sim \overline{\ddot{M}_{22}}$.}
        \label{Case1}
\end{figure}

\begin{figure}
     \centering
     \begin{subfigure}[b]{0.49\textwidth}
         \centering
         \includegraphics[width=\textwidth]{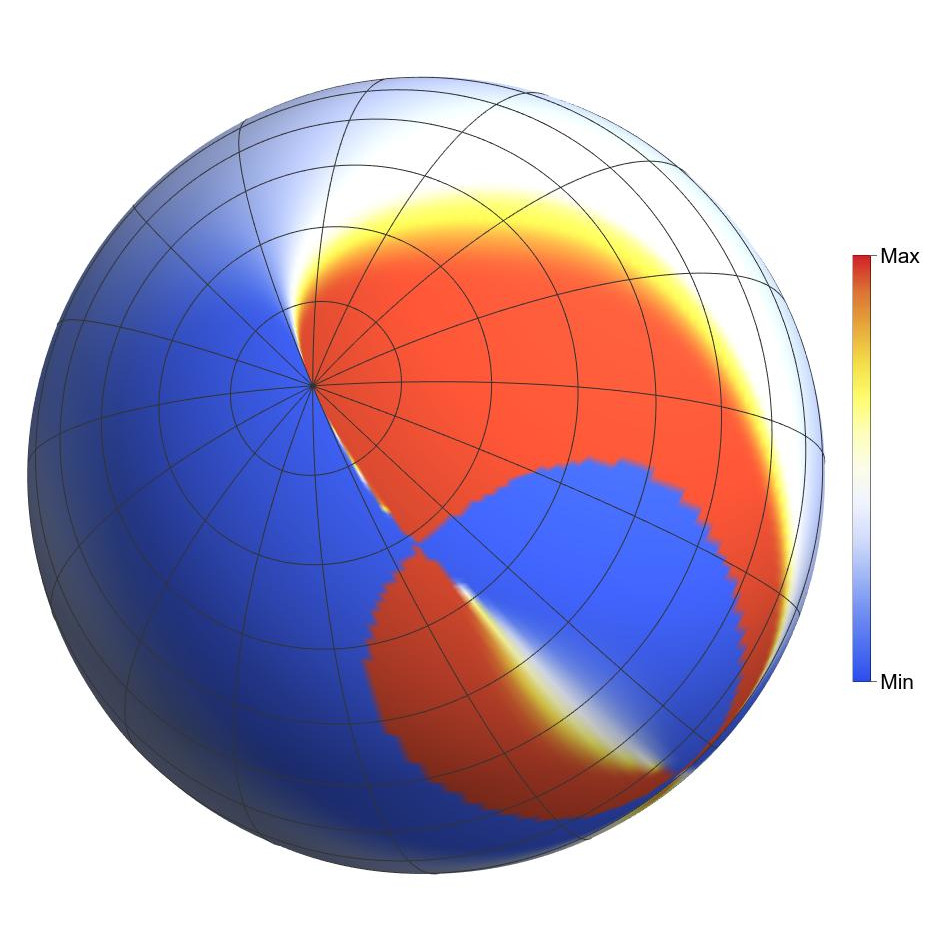}
         \caption{$\frac{\Delta {\cal C}_v}{{\cal C}_v}$, view from the north pole.}
         \label{cv2}
     \end{subfigure}
     \begin{subfigure}[b]{0.49\textwidth}
         \centering
         \includegraphics[width=\textwidth]{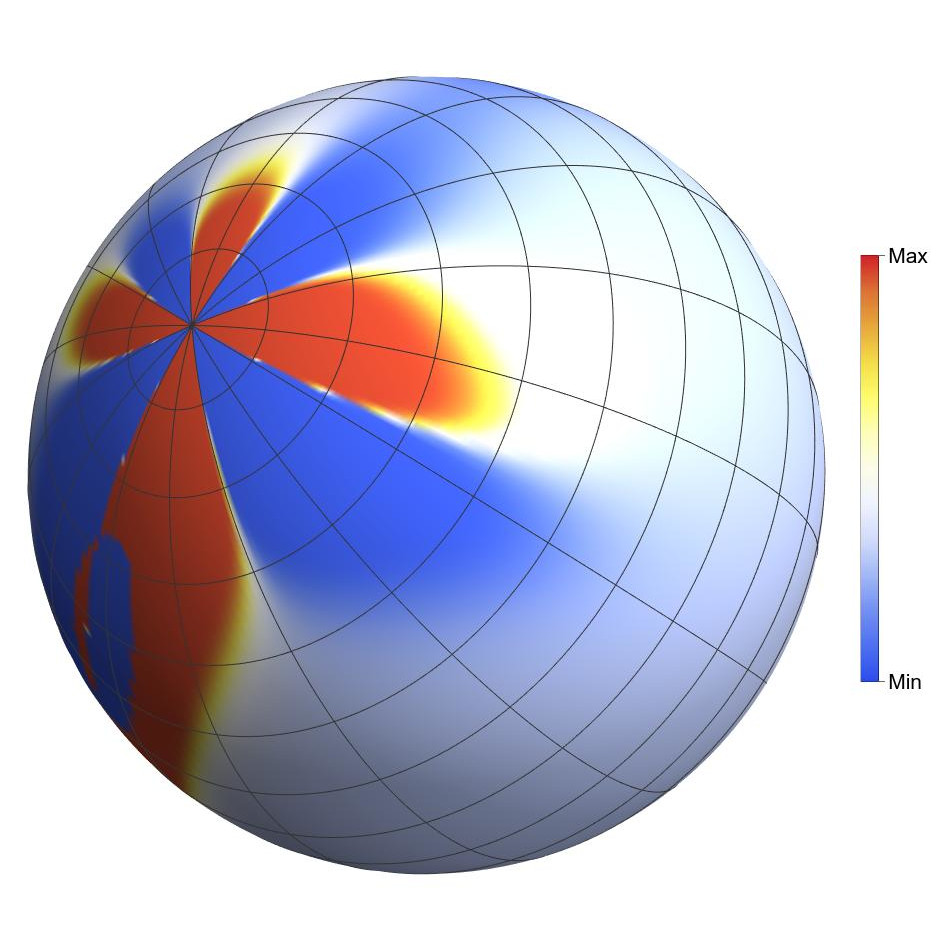}
         \caption{$\frac{\Delta {\cal C}_v}{{\cal C}_v}$, view from the south pole.}
         \label{cv2t}
         \end{subfigure}
        \caption{Dependence of $\frac{\Delta {\cal C}_v}{{\cal C}_v}$ on the direction of incoming gravitational pulse for the Case $\overline{\ddot{M}_{13}}\sim \overline{\ddot{M}_{23}}\sim \overline{\ddot{M}_{33}} \sim \overline{\ddot{M}_{22}}\ll \overline{\ddot{M}_{11}}$.}
        \label{Case2}
\end{figure}

As a result an observer falling with our box can detect the direction of incoming gravitational pulse and also its magnitude (since the absolute value of $\Delta\Phi$ also depends on the values of the mass moments) by measuring relative changes in thermodynamic quantities, e.g. $\frac{\Delta {\cal C}_v}{{\cal C}_v}$. Figures (\ref{Case1}) and (\ref{Case2}) illustrate the this angular dependence for the two cases mentioned before. An important property is that the value of $\frac{\Delta {\cal C}_v}{{\cal C}_v}$ is highly sensetive to the orientation of the incoming wave with respect to the direction of magnetic field.

Similar graphs for other types of gravitational wave sources can be obtained. Also for any other thermodynamic quantity, $Q$, graphs of $\frac{\Delta Q}{Q}$ can be used for determining the direction and magnitude of the incoming gravitational pulse.

%%%%%%%%%%%%%%%%%%%%%%%%%%%%%%%%%%%% Section IV
\section{Concluding Remarks} \label{SecIV}
The gyroscopic memory has recently been introduced as the precession of a gyroscope with respect to the distant galaxies when a localized  gravitational wave pulse passes a freely falling gyroscope \cite{Seraj1}. In this paper, we studied the thermodynamic properties of a canonical ensemble of such gyroscopes just after the passage of the gravitational wave.

We considered a freely falling box of classical paramagnetic matter in a magnetic field, and far from the gravitational wave source. As a pulse of gravitational wave passes through the box and disturb the falling trajectory, the partition function of this ensemble would experience a change because of change in the Hamiltonian of the system. We saw that this change of the thermodynamic properties is proportional to the precession memory effect factor $\Delta\Phi$, Eq. \eqref{PF}. 

All the thermodynamic quantities of the paramagnetic matter in the box, would experience such a change. As a result each one can be used as an observing apparatus for the measurement of the direction and magnitude of the incoming gravitational wave.
For two special cases, the dependence of $\frac{\Delta{\cal C}_v}{ {\cal C}_v}$ on the direction is plotted.

In order to see if such an apparatus can practically be used as a gravitational wave detector, we explore the effect of gravitational waves emitted by a binary system of neutron stars (BNS) on the specific heat of a box of paramagnetic matter. As it is clear from equations \eqref{memory}, \eqref{A11} and \eqref{A22}, $\Delta \Phi$ is proportional to the second time derivative of the second mass moment, $\ddot{M_{ij}}$. For a BNS, in its spiral phase, \cite{Blanchet1995} we have
\begin{multline}
\ddot{M}_{ij}(t) \propto 2\nu M \left(\alpha^{ij} - \frac{G M}{r^3} x^{ij} + \gamma \frac{G M}{r^3} \frac{x^{ij}}{42} (149-69\nu) - \frac{\gamma}{42}\alpha^{ij} (23-27\nu) \right. \\
\left. + \frac{\gamma^2}{1512}  x^{ij}\frac{G M}{r^3} \left(-7043+7837\nu-3703\nu^2 \right) +\frac{\gamma^2}{1512} \alpha^{ij}\left(-4513-19591\nu+1219\nu^2 \right) \right) \ ,
\label{MassMoment}
\end{multline}
where $m_1$ and $m_2$ are the masses of the components of the BNS, with rotation period $\tau$ and 
\begin{align}
& M = m_1+m_2     && \nu = \dfrac{m_1 m_2}{M^2}   && r = (\dfrac{G M \tau^2}{4\pi^2})^{1/3} \ \nonumber \\
& \alpha^{ij} = v^i v^j = \dv{x^i}{t} \dv{x^j}{t} \sim v^2    && x^{ij} = x^ix^j \sim (2r)^2   && \gamma = \dfrac{G M}{c^2 r} \ .
\end{align}
where $v$ is the linear velocity of each of the stars. The change in the specific heat is not only dependent on $\Delta \Phi$, but also on the local parameters of the paramagnetic matter, like the magnetic moment of the atoms of the ensemble, $\mu$, and the local magnetic field $B$. 

In order to have an estimation of the memory effect for the specific heat, consider Tungsten or Sodium as a typical paramagnetic matter with magnetic susceptibility of $6.8\times 10^{-5}$ and $0.72\times 10^{-5}$, respectively. Also consider that this paramagnetic matter is placed within a magnetic field with magnitude of about $100-1000$ Gauss, say. For the source of the gravitational wave we choose a BNS with mass of each component in the allowed range of $1.1 M_{\odot} < m < 1.7 M_{\odot}$, and for BNS\footnote{The numerical data are from \cite{numerical-data}.} the rotating period is something between $0.1 \text{ day}$ and $45 \text{ days}$. To have an estimation for $\Delta\Phi$, we also need the distance to the BNS and the duration of the gravitational pulse (see equation (\ref{memory})). Assuming the distance to the BNS be of order $10\text{ Mpc}$, the value of $\dfrac{\Delta C_v}{C_v}$ per duration time of the pulse\footnote{As we have approximated the integral of $\ddot{M}_{ij}$ over $u$ in equation \eqref{memory} as $\overline{\ddot{M}_{ij}}\times T$ (with $T$ the pulse duration) and as $\dfrac{\Delta C_v}{C_v}$ is linear in $\Delta\Phi$ and thus in $T$, we present here the value of $\dfrac{\Delta C_v}{C_v}$ for a pulse duration of one second. Therefore the  value of $\dfrac{\Delta C_v}{C_v}$ is actually the obtained value multiplied by the gravitational pulse duration time.} can be estimated from equations (\ref{memory}), (\ref{SH}) and (\ref{MassMoment}). 

\begin{figure}[ht]
\centering
\includegraphics[width=0.6\textwidth]{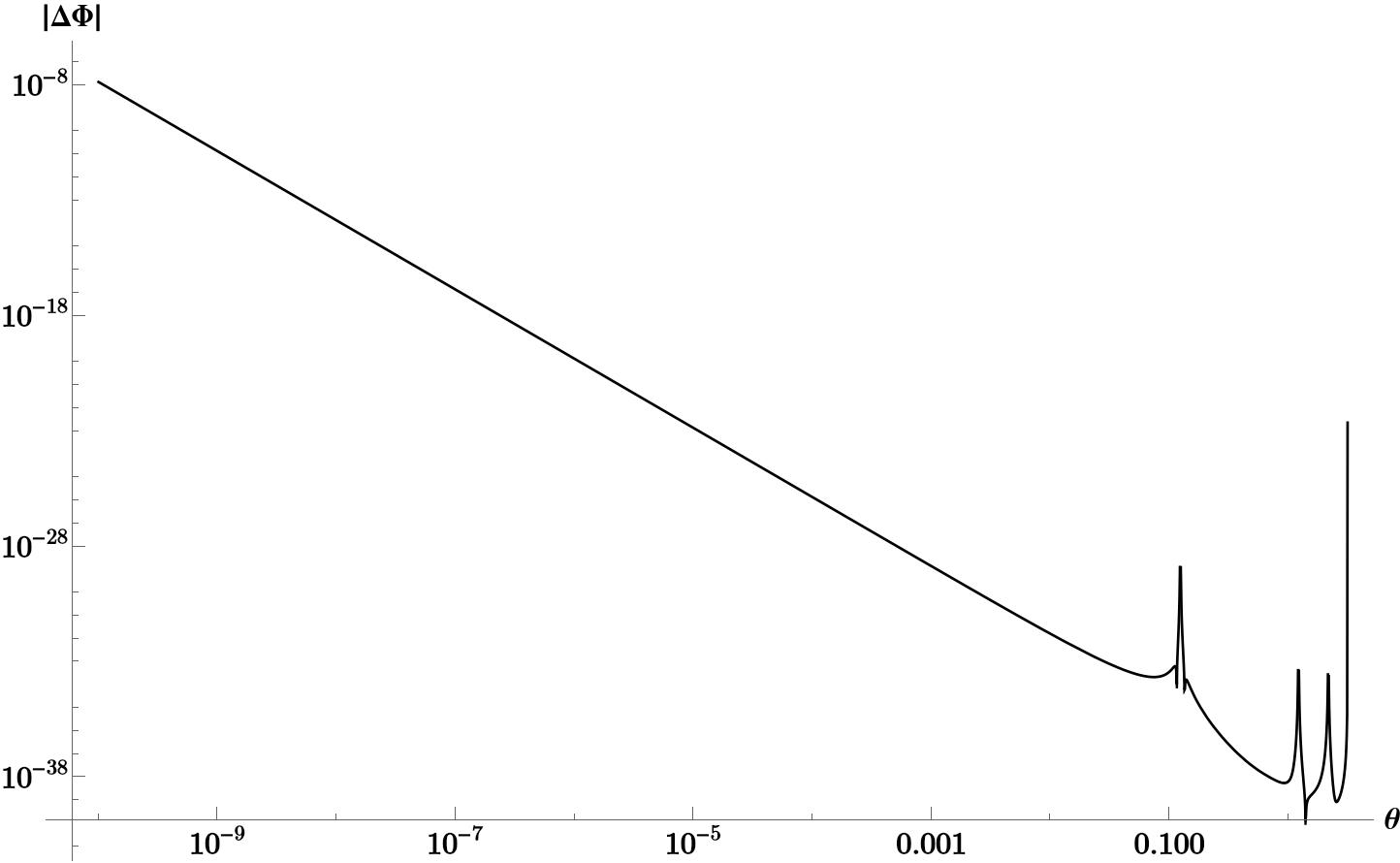}
\caption{Dependence of $\Delta\Phi$ on the angle between the incoming wave and the magnetic field for a typical BNS mentioned in the text, with $m=1.1 M_{\odot}$ and $\tau= 0.1 \text{ day}$.}
\label{del}
\end{figure}

\begin{figure}[h]
     \centering
     \begin{subfigure}[b]{0.49\textwidth}
         \centering
         \includegraphics[width=\textwidth]{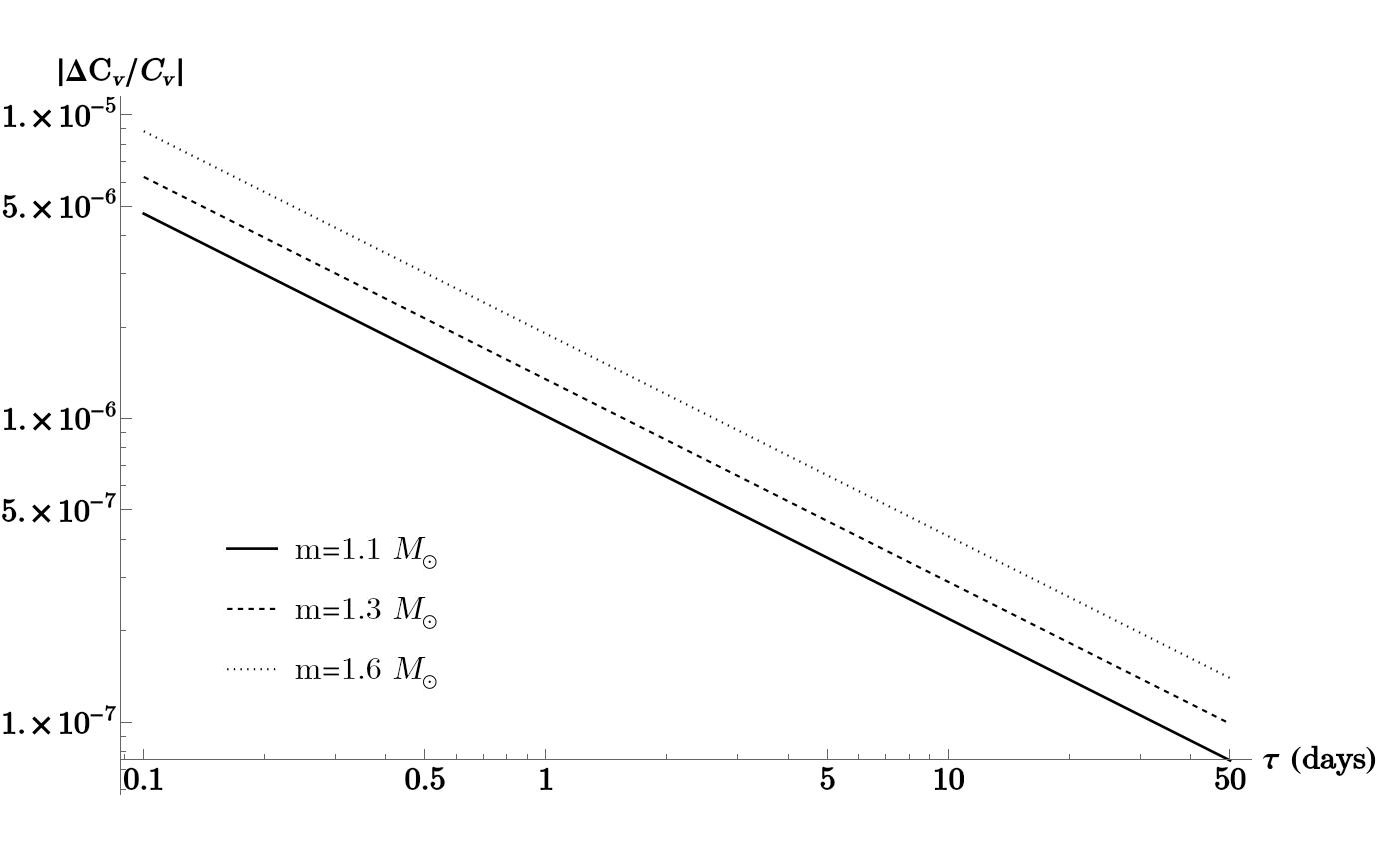}
         \caption{$\vert\frac{\Delta {\cal C}_v}{{\cal C}_v}\vert$ at $\theta=0$ and $\phi=\pi/10$.}
         \label{N1}
     \end{subfigure}
     \begin{subfigure}[b]{0.49\textwidth}
         \centering
         \includegraphics[width=\textwidth]{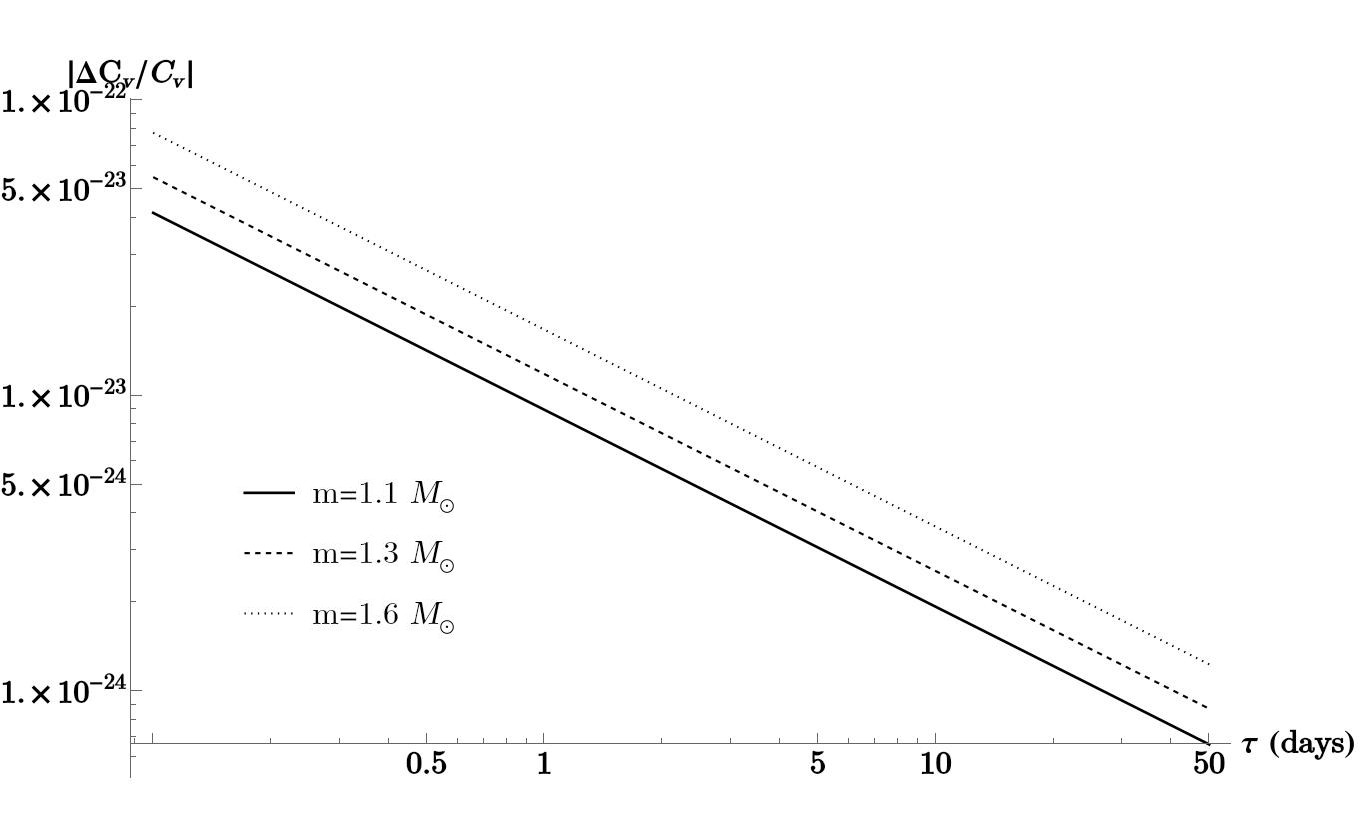}
         \caption{$\vert\frac{\Delta {\cal C}_v}{{\cal C}_v}\vert$ at $\theta=1''$ and $\phi=\pi/10$.}
         \label{N2}
         \end{subfigure}
        \caption{The heat capacity memory effect for Tungsten as a function of rotating period and different mass of the BNS. Note the dramatic change in the result by only one arcsec of change in the direction. The Tungsten paramagnetic matter is placed within $B=100\text{ Gauss}$ and is located about $10\text{ Mpc}$ far from the BNC. The temperature of the matter is assumed to be $500K$.}
        \label{NumericalAnalysis}
\end{figure}

Note that since the value of $\Delta\Phi$ is highly dependent on the angles, the drawback of such a measurement is its sensitivity to the exact alignment of local magnetic field and the direction of the incoming wave. In Fig. \ref{del}, one can see how $\Delta\Phi$ is a fast changing function. Also in Fig. \ref{NumericalAnalysis} the behavior of memory effect of Tungsten for the above mentioned conditions and for two directions differing only by one arcsec is shown. Note the dramatic change in the result by only one arcsec of change in the direction.

Therefore values of about $10^{-7}-10^{-6}$ for $\dfrac{\Delta C_v}{C_v}$ per pulse duration time are accessible. As a result if the experimenter changes the temperature of one kilogram of Tungsten by $10$ degrees, the memory effect causes that the necessary energy increases by about $10 \mu J$--$10 mJ$ per pulse duration time. This is quiet measurable, provided that the experimenter is able to shield the apparatus from other effects successfully and can align the apparatus exactly.

Another important fact is that there are contribution from two terms, the spin- memory and electric-magnetic duality, in $\Delta\Phi$. It becomes evident that for all of the range of $(\theta,\phi)$, the primary contribution to $\dfrac{\Delta C_v}{C_v}$ comes from the former. For our example of BNS, contribution from electric-magnetic duality term is always below $\sim 20\%$, as it can be seen in Fig. \ref{Contribution}.

\begin{figure}[ht]
 \label{Contribution}
     \centering
     \begin{subfigure}[b]{0.48\textwidth}
         \centering
         \includegraphics[width=\textwidth]{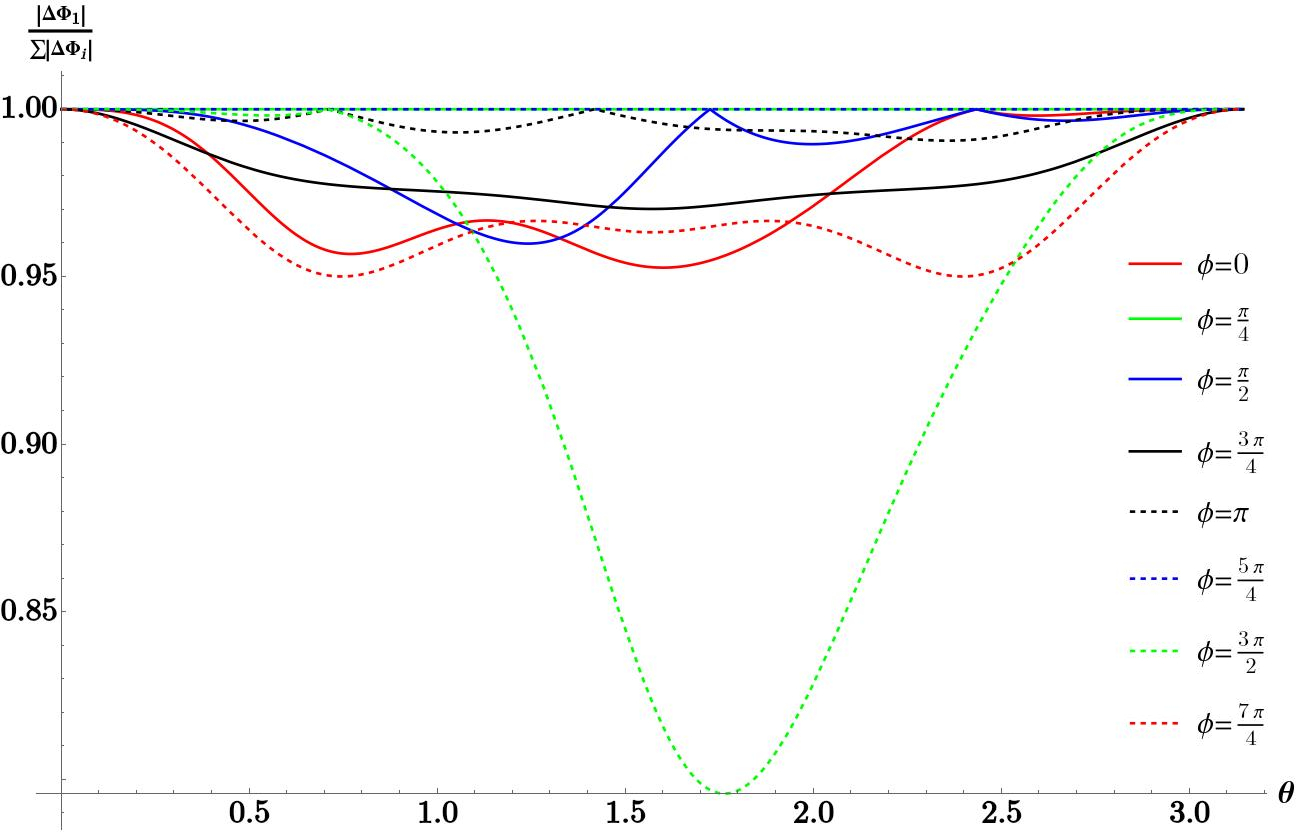}
         \caption{$\Delta\Phi_1$}
         \label{Contribution1}
     \end{subfigure}
     \begin{subfigure}[b]{0.48\textwidth}
         \centering
         \includegraphics[width=\textwidth]{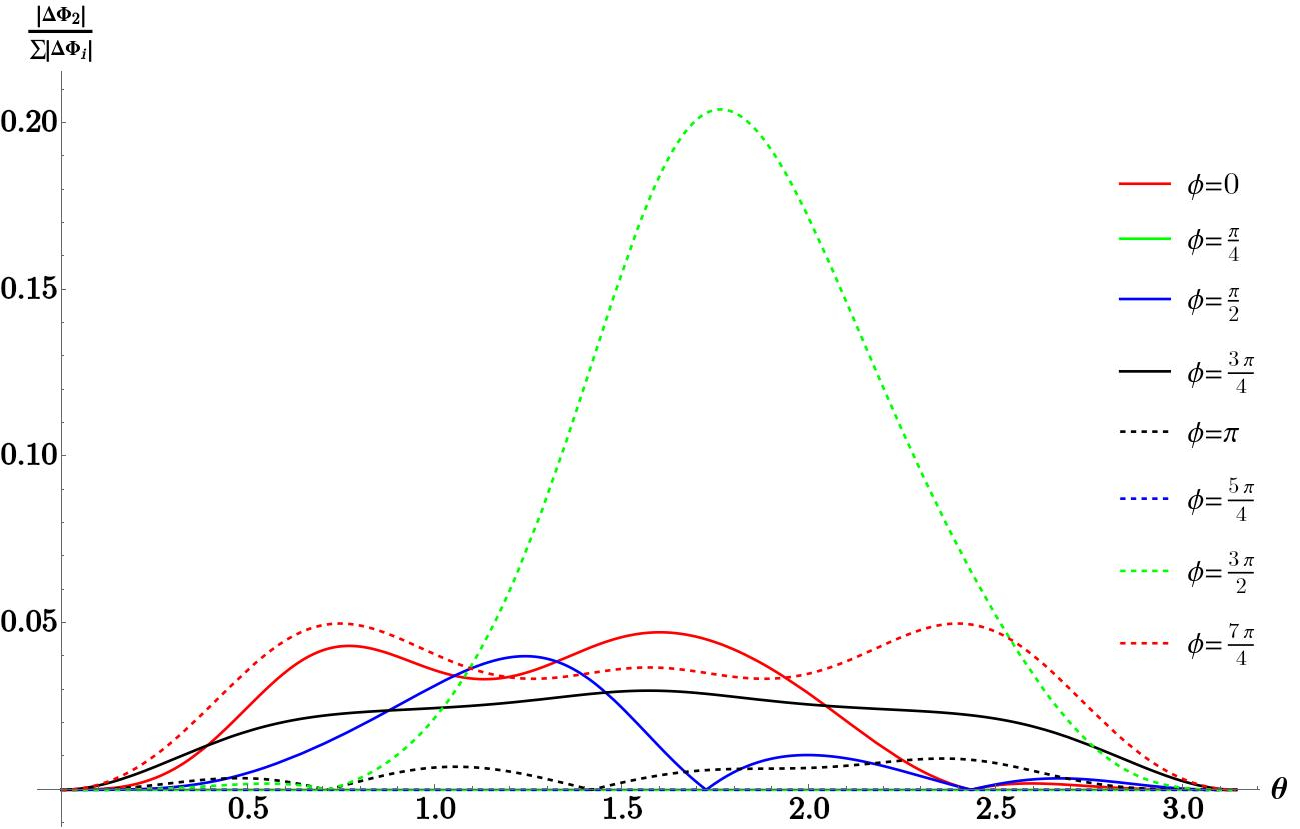}
         \caption{$\Delta\Phi_2$}
         \label{Contribution2}
         \end{subfigure}
     \caption{Contribution of the spin-memory effect $(\Delta\Phi_1)$ and the electric-magnetic duality memory $(\Delta\Phi_2)$ to the total $\Delta\Phi = \sum_{i=1}^2 \abs{\Delta\Phi_i}$.}
         \label{Contribution}
\end{figure}

The measurability of such an effect is even better for binary black hole systems. For example, for GW150914 with total mass of about $65 M_\odot$, distance of $410\text{ Mpc}$, orbital frequency of about $75\text{ Hz}$, and signal duration of about $0.2 \text{ sec}$\cite{GW-BH1,GW-BH2}, one is dealing with $\dfrac{\Delta C_v}{C_v}$ of order $10^{-3}$ when the magnetic filed is exactly aligned with the incoming wave.
  
At the end, it should be noted that such an effect, the GravoThermo memory effect, does not conflict with the quantum nature of the paramagnetic moments. In fact, for quantum paramagnetic matter the same calculations can be done in a similar way and the results are not expected to be very different. 
This is because of the fact that for the quantum case, instead of the integration over whole the configuration space (orientation of the spin) we only have to sum over the eigenvalues of the Hamiltonian. Although this changes the result but the fact that it is still proportional to $\Delta\Phi$ leads to some $\dfrac{\Delta C_v}{C_v}$ also  proportional to $\Delta\Phi$ and of the same order. For the quantum case, the spin is intrinsic and the precession memory effect is the result of the first term of \eqref{memory}.

%%%%%%%%%%%%%%%%%%%%%%%%%%%%%%%%%%%% Appendix
\appendix
\renewcommand{\thesection}{Appendix}
\section{Bondi Space-time} \label{App. A}
The asymptotic symmetries of Bondi space-time, provides a comprehensive approach for describing all of the memory effects, specially the gyroscope one. This metric, revealed the existence of the symmetry group, Bondi--Metzner--Sachs (BMS) group, at null infinity, which is the extended version of the Poincare group, supertranslation \cite{Sachs, Flangan}. The memory effects are described properly with the Noether charges of the group\footnote{Besides, the fundamental symmetries of gravity which clearly follow in BMS approaches may open the way to the quantization of gravity\cite{Strominger1, Strominger2, Mao}.} \cite{Strominger1, Freidel, Kol, Godazgar} .

The retarded Bondi coordinates $x^{\alpha} = (u,r, \chi^{1},\chi^{2})$ are based on a family of outgoing null hypersurfaces $u=const$ where  $u \equiv t-r$ is the retarded time, $r$ varies along the null rays and the $\chi^{A}=(\theta,\phi)$ are  angular coordinates which are constant along the null rays.

Near the future null infinity the metric has the form 
\begin{equation}
\dd{s}^2 = -Ue^{2\tilde{\beta}} \dd{u^2} - 2e^{2\tilde{\beta}} \dd{u} \dd{r} +r^2\gamma_{AB}\left(\dd{\chi^A}-\mathcal{U}^A\dd{u}\right)\left(\dd{\chi^B}-\mathcal{U}^B\dd{u}\right)
\end{equation}
where $A,B=1,2$ and $U, \tilde{\beta}, \mathcal{U}^{A}$ and $\gamma_{AB}$ are functions of coordinates $(u,r,\chi^A)$.

The Bondi gauge satisfies four conditions 
\begin{equation}
g_{rr}=g_{rA}=0 \quad \quad, \quad \quad \partial_r \text{det}(r^{-2} g_{AB}) = 0 \ .
\end{equation}
Since we are at the future null infinity, expanding the metric as series in $1/r$ simplifies the relations \cite{Pasterski, Sachs, Flangan, Bondi, Sachs2, Barnich}. The order at which the various expansions start can be deduced from the covariant definition of the asymptotic flatness at future null infinity \cite{Wald}. They are 
\begin{align}
& \tilde{\beta} = \frac{\tilde{\beta}_0}{r} + \frac{\tilde{\beta}_1}{r^2} + \frac{\tilde{\beta}_2}{r^3} + \mathcal{O}(r^{-4}) \\
& U = 1 - \frac{2 M_{\mathcal{B}}}{r} - \frac{2 \mathcal{M}}{r^2} + \mathcal{O}(r^{-3}) \\
& \gamma_{AB} = h_{AB} + \frac{1}{r} C_{AB} + \frac{1}{r^2} D_{AB} + \mathcal{O}(r^{-4}) \\
& \mathcal{U}^{A} = \frac{1}{r^2} U^A + \frac{1}{r^3} \left[-\frac{2}{3} L_{\mathcal{B}}^A + \frac{1}{16} \nabla^A(C_{BC}C^{BC}) +\frac{1}{2} C^{AB}\nabla^C C_{BC} \right]  + \mathcal{O}(r^{-4}) \ .
\end{align}
The coefficients on the right hand sides of these relations (i.e. $\tilde{\beta}_{i=0,1,2}, M_{\cal{B}}, \mathcal{M}, C_{AB},  L_{\mathcal{B}}^A$ and $D_{AB}$) are functions of $(u,\chi^A)$. The leading-order coefficients $M_{\mathcal{B}}$, $L_{\mathcal{B}}^A$ and $C_{AB}$ are known as Bondi mass aspect, Bondi angular momentum aspect and Bondi shear, respectively. Except $M_{(\mathcal{B})}$ and $L_{(\mathcal{B})}^A$, all the others are functions of the Bondi shear $C_{AB}$ which describes the time dependence of the gravitational radiation on the gravitational wave source \cite{Seraj4}. For $\chi^A=(\theta,\phi)$ the metric $h_{AB}$ on the unit 2-sphere, is $\text{Diag} (1,\sin^2\theta)$. The capital Latin indices $A$ and $B$, are raised and lowered with this metric. Moreover, the $\nabla_A$ is covariant derivative associated with it. Note that the Bondi shear asymptotically behaves as
\begin{equation} \label{Shear}
 C_{AB} \simeq \lim\limits_{r\to\infty} r \left(\gamma_{AB}-h_{AB} \right).
\end{equation}

\end{document}